\documentclass[11pt,twoside]{article}

\usepackage{asp2006}
\usepackage{epsf}
\usepackage{lscape}
\usepackage{graphicx}

\begin{document}
\title{V2467 Cygni as possible intermediate polar}   
\author{Swierczynski E., Mikolajewski M., Tomov T., Ragan E., Galan C., Karska, A., Wychudzki P., Wiecek M., Cikala M. and Lewandowski M.}   
\affil{Centre for Astronomy, Nicolaus Copernicus University, \\ 87-100 Torun, ul. Gagarina 11, Poland }    

\begin{abstract} 
We present the results of unfiltered and $\mathrm{UBVR_cI_c}$ band CCD photometry fast nova V2467Cyg.
Our analysis of the data gives two distinct frequencies corresponding to periods of
~3.8h and ~35 min. The observed light curve of V2467 Cyg is typical for an intermediate polar.
\end{abstract}

\section{Introduction}

Classical novae belong to wide class of cataclysmic variables characterized by
such commonly features as accretion from the Roche lobe filling star to its
compact companion (most often white dwarf) and by outbursts connected with
rapid rise in brightness. Location of accretion material depends on angular momentum
distribution and on magnetic field of primary. In most cases, stream of matter
forms keplerian disc around white dwarf before fall onto its surface. In some systems
with the magnetic white dwarf, called intermediate polar, field is strong enough
($\mathrm{10^5 G<B_{0}<10^8 G}$) to allow material at the outer edge of disc to become
magnetically controlled. Accretion can be channeled
by bipolar magnetic field on small fraction of primary surface near poles. Both hot regions are
sources of anisotropy of emission observed during rotation of white dwarf in X-ray and optical.

Here we present observations of the fast nova V2467 Cygni, which was discovered by Ahikigo
Tago in March 2007 at 7.4 mag \cite{nakano}. About twenty days after outburst, extremely
strong OI line at 8446$\mathrm{\AA}$ was presented in spectra as result of overabundance of oxygen
\cite{tomov}. During the transition phase, which was started in April, six quasiperiodic
oscillations with period from 19 to 25 days and amplitude about 0.7 mag were observed \cite{kato}.
 One year after this, \cite{ness} reported
that V2467 Cygni is a soft X-ray source and blackbody fitted to its spectra gave parameter kT=34eV.

\section{Observations and Analysis}

We obtained our photometric data using the Cassegrain 60cm telescope with SBIG STL-1001 camera located in Piwnice
Observatory near Torun. The data was collected during two observational seasons from 12 April 2007 to 28 May 2007 and
from 25 june to 23 October 2008 and include mesurements taken with $\mathrm{UBVR_cI_c}$ filters to construct long-term light curve. 
At the same time, we carried out thirty one-night monitorings to find short time scale variability 
mainly in V and $\mathrm{R_c}$ passband in 2007 and of white light in 2008. First search results
was presented in \cite{tomov} and are agreeable with the one presented below.

\subsection{2007 data}

Transition phase of this nova started about one month after outburst when nova faded about 4 magnitudes. 
In this time,
we monitored this star about 30 hours in $\mathrm{R_c}$ and 24 hours in V passband. We analysed data
in two filters separately with Fourier methods.
The best fitted periods were 3.46.5 min and 3.40.5 min in $\mathrm{R_c}$ and V respectively.
\setcounter{figure}{1}
\begin{figure}[!ht]
\begin{center}
\includegraphics[scale=0.3, angle = 270]{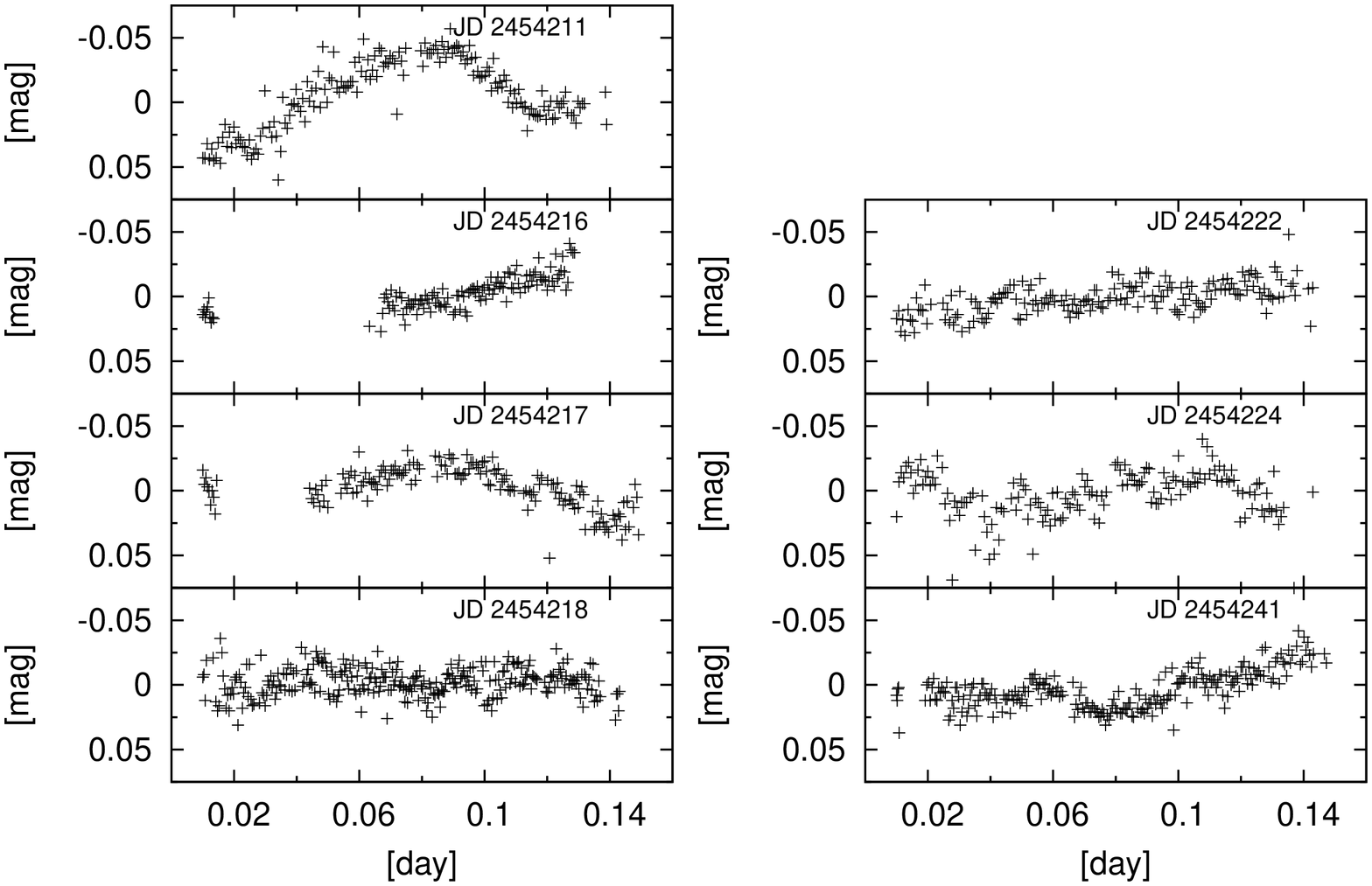}
\end{center}
\caption{All long monitorings  of nova in 2007 obtained in V passband}
\end{figure}

\subsection{2008 data}

In 2008, over year after outburst nova was too faint to observe with our telescope and with filters,
so we 
observed white light. In all, we collected 22 hours of observations which showed
two significal variabilities. We confirm that longer variability, about 3.45.5 min, was observed also in 2007.
\setcounter{figure}{3}
\begin{figure}[!ht]
\begin{center}
\includegraphics[scale=0.3, angle = 270]{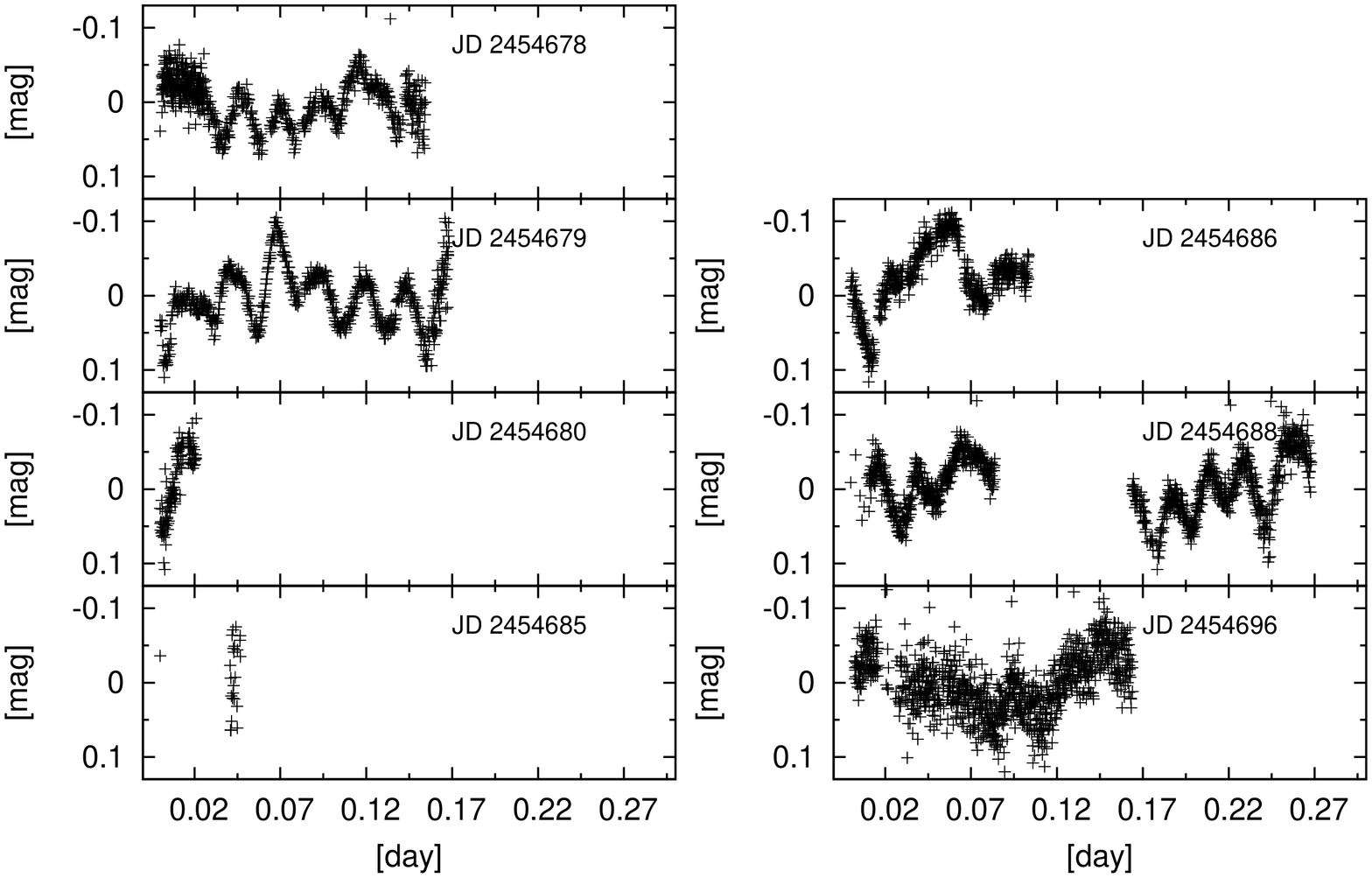}
\end{center}
\caption{All long monitorings  of white light of nova in 2008.}
\end{figure}

Our attention was concentrated on stable changes with period
about 40 minutes and amplitude about 0.15 mag. 
We  don't have any evindence for exsist changes with this period
in 2007 data.
This period is too short to connect its with orbital motion. We used timing methods on maximas
to check if this changes are coherent. On base of this measurements we constructed O-C diagram and computed
ephemeris $ \mathrm{HJD_{max} = 2454678.471(6) + 0.024062(2)*E}$.
\setcounter{figure}{5}
\begin{figure}[!ht]
\begin{center}
\includegraphics[scale=0.3, angle = 270]{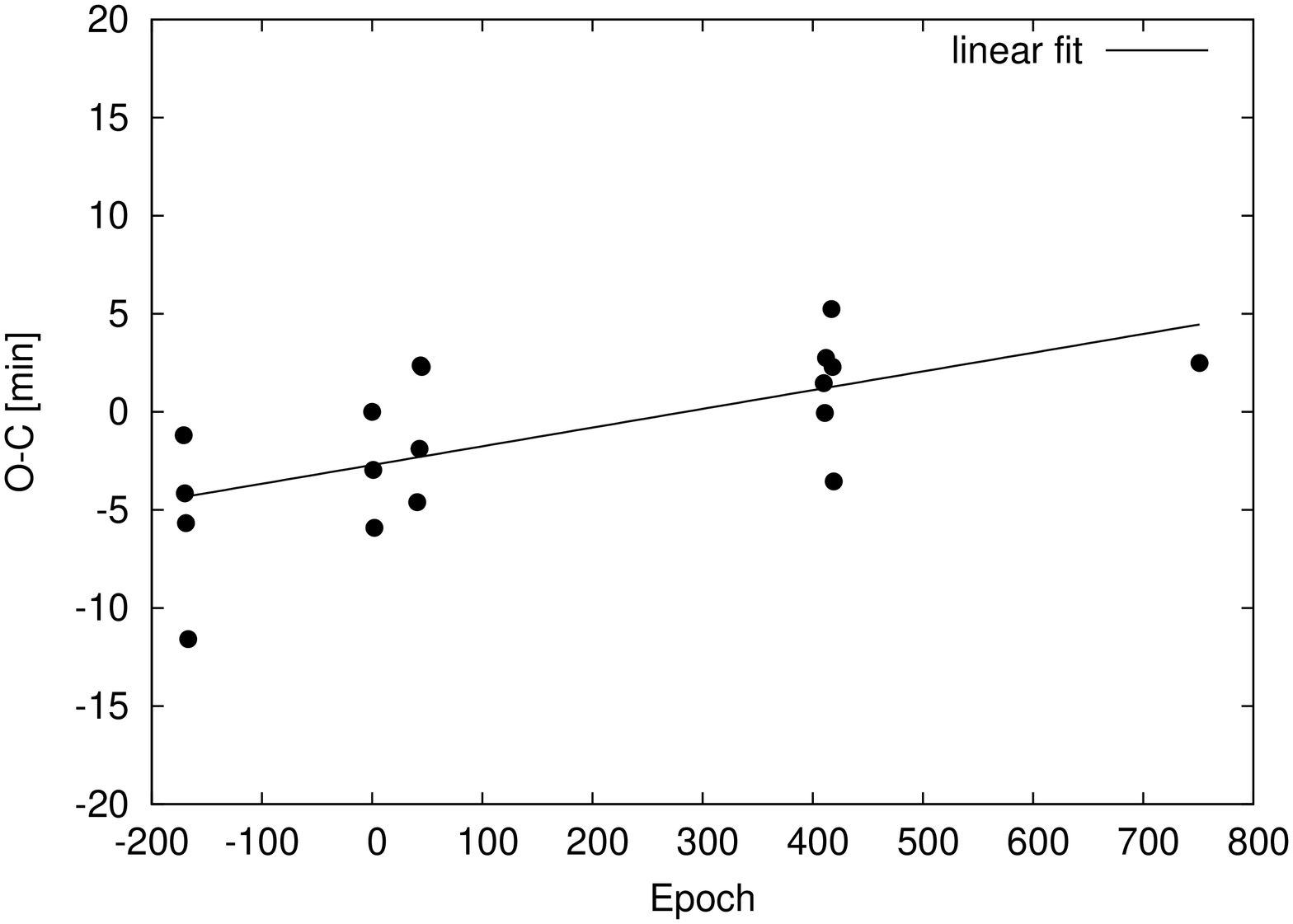}
\end{center}
\caption{O-C diagram for period 34.5 min}
\end{figure}
\setcounter{figure}{6}
\begin{figure}[!ht]
\begin{center}
\includegraphics[scale=0.3, angle = 270]{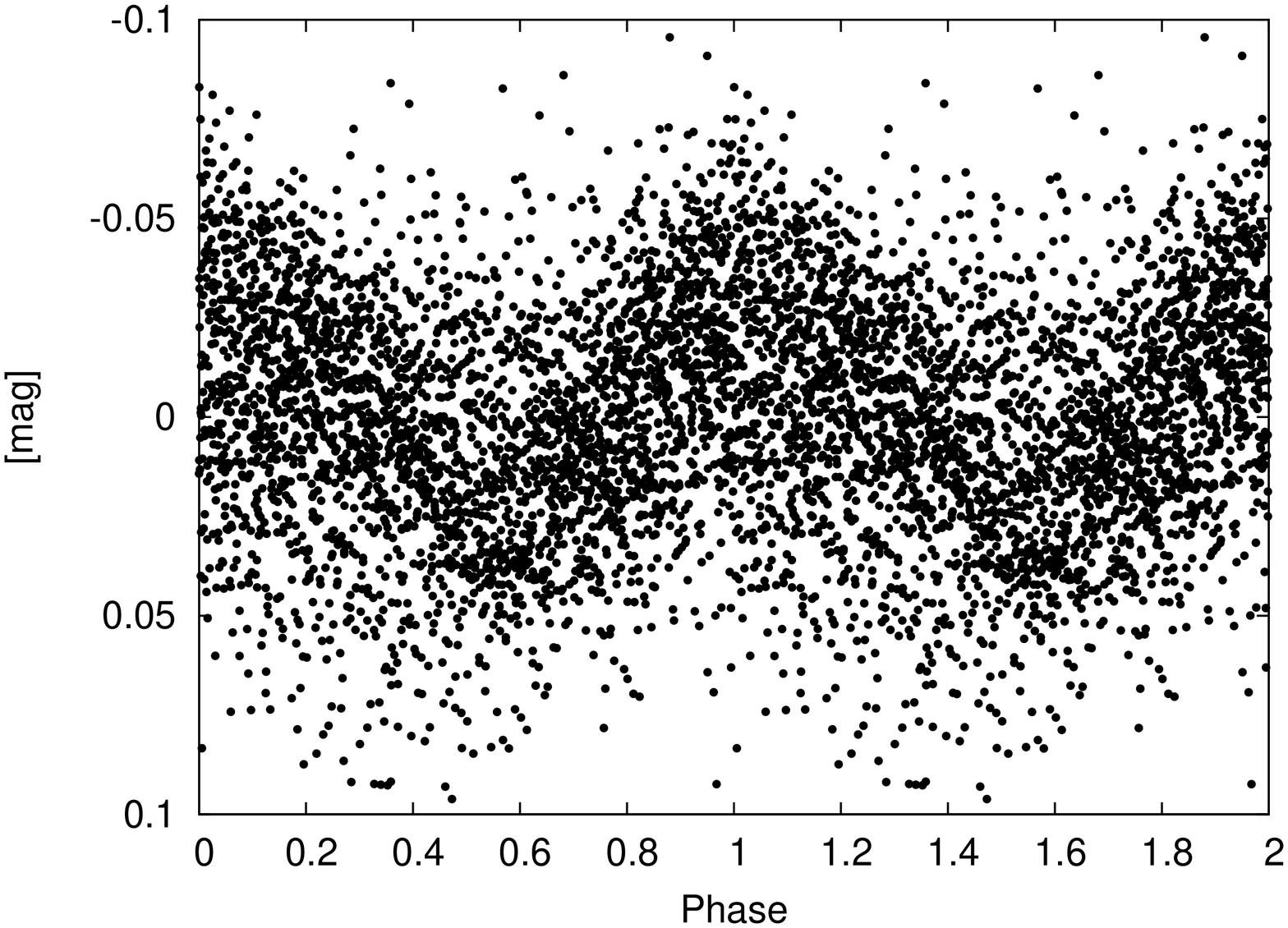}
\end{center}
\caption{Data phased with period from O-C diagram after prewhitening.}
\end{figure}

\subsection{X-ray data}

We completed X-ray data obtained in September 2008 using Swift satellite include 1500 second
of observation. Eight points, each include information about several minutes of expositions, 
were phased with ephemeris showed above. We attached to single points also errorbar connected with
exposition time (x-axis) and range from maximum and minimum counts rate (y-axis).    

\setcounter{figure}{7}
\begin{figure}[!ht]
\begin{center}
\includegraphics[scale=0.29, angle = 270]{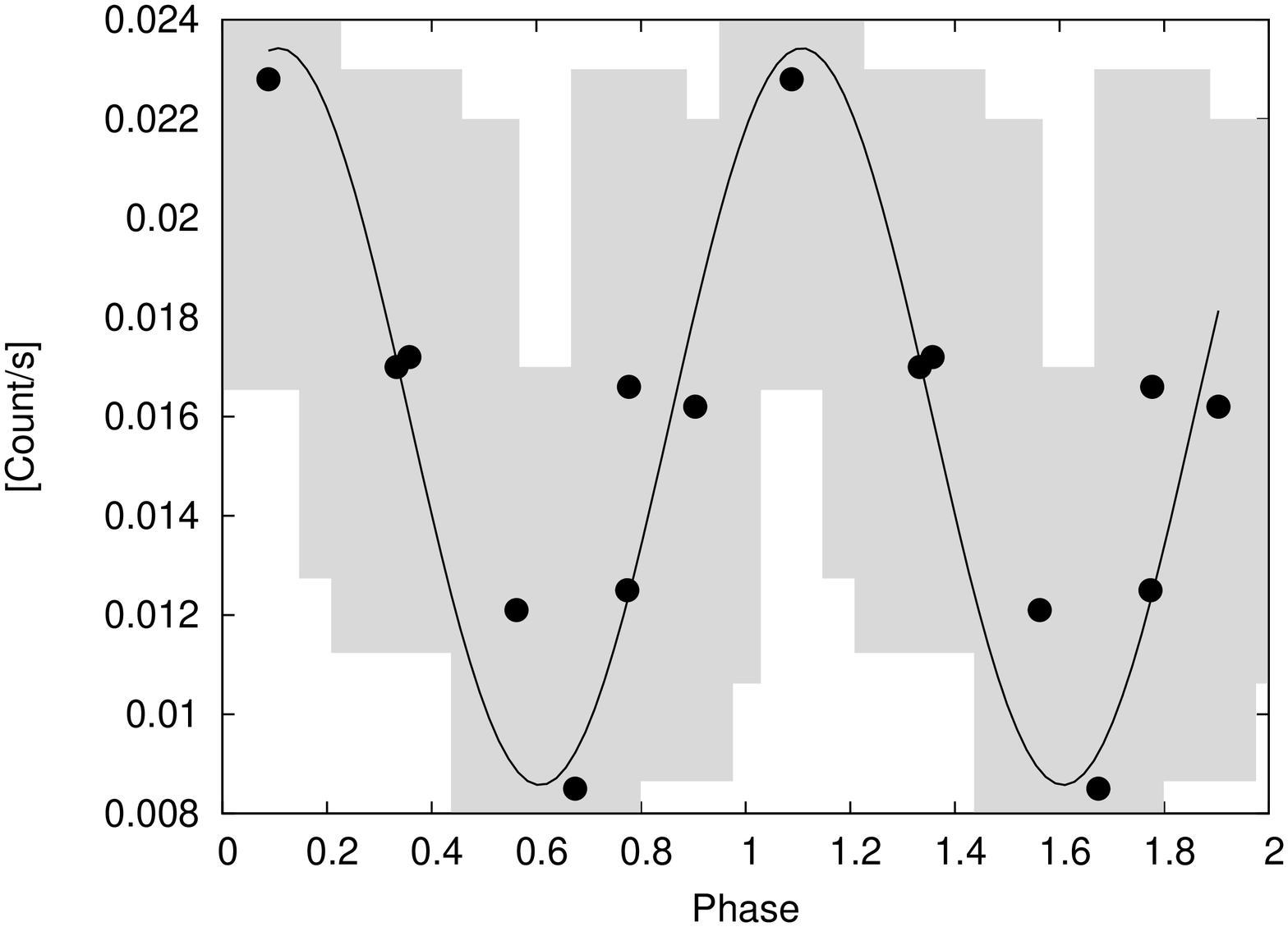}
\end{center}
\caption{Data in X-ray phased with ephemeris computed for short period}
\end{figure}
\section{Conclusion}
We found short coherent period in optical light curve and confirm that there changes are also visible in X-ray.
Our colour light curve show that amplitude of short variations increase in longer wavelength, so cannot be connected
with disc instabillity or with pulsations of white dwarf.
 We can explain it with intermediate polar model
when magnetic field have influence on accretion near white dwarf. 
\section{Acknowledgments}
This work was supported by the Polish MNiSW Grant N203 018 32/2338.






\textsl{M. Friedjung}: How much had the nova faded, when you saw the 3h periodicity? Had the optically thick wind stopped by then?
Did nova V2467 have oscillations several magnitudes below optical maximum?

\textsl{E. Swierczynski}: Nova had faded quickly about 4 magnitudes to the point when we could observed radiation from the star itself.
In the same time, V2467 Cygni was entering brightness oscilations during transition phase with period from 19 to 25 days.

\textsl{S. Katajainen}: Did you calculated beat period using possible spin of 40 min and $ \mathrm{P_{orb}} $ = 3h48min? Any signs of beat period
in fourier analysis? What about Balmer lines in spectrum? 

\textsl{E. Swierczynski}: In fact, we observed short variability as superposition of orbital
motion and rotaion of white dwarf. In periodogram we can see sign of beat period but not clearly. 
We have not any spectra obtainted in this time.
\end{document}